\documentclass[preprint,12pt]{elsarticle}
\usepackage{graphicx}
\usepackage{amssymb}
\usepackage{bm}
\usepackage{subfigure}
\graphicspath {{./}}
\journal{J. Nucl. Mater.}

\begin{document}

\begin{frontmatter}

\title{Hydrogen influence on generalized stacking fault of Zr \{0001\} plane: 
 a first-principles calculation    study   }

\author{Songjun Hou$^1$, Huaping Lei$^1$,  Zhi Zeng$^{1,2}$*}

\address{*Corresponding Author: zzeng@theory.issp.ac.cn \\
$^1$Key Laboratory for Materials Physics, Institute of Solid State
Physics, \\ Chinese Academy of Sciences, Hefei, 230031, China \\
$^2$Department of Physics, University of Science and Technology of China, \\
Hefei, 230026, China \\
}
\begin{abstract}
  The influences of hydrogen on the generalized stacking fault (GSF) energies of the 
  basal plane along the $<10\bar{1}0>$ and $<11\bar{2}0>$ directions in hcp Zr were investigated 
   using the first-principles calculation method. 
  The modifications of the GSF energies were studied with respect to 
  the different distances of H atoms away from the slip plane and hydrogen content there. 
  The calculation results have shown that the GSF energies along the $<10\bar{1}0>$ direction 
  drastically reduce when H atoms locate nearby the slip plane. 
  But H atoms slightly decrease the GSF barrier for the $<11\bar{2}0>$ slipping case. 
  Meanwhile, with the increase of hydrogen density around the slip plane, 
  the GSF energies along both the two shift directions further reduced greatly. 
  The physical origin of the reduction of GSF energies due to the existence of hydrogen atoms 
  in Zr was analyzed based on the Bader charge method. 
  It is interpreted by the Coulomb repulsion of the Zr atoms beside the slip plane 
  due to the charge transfers from Zr to H .

\end{abstract}

\begin{keyword}
First principles   \sep hydrogen  \sep Zirconium  \sep stacking fault energy \sep Bader charge
\end{keyword}

\end{frontmatter}
\section{Introduction}
The generalized stacking fault (GSF), as discussed by Vitek\cite{Vitek1968}, 
is often correlated with the dislocations mobility 
and plasticity of materials. The mechanical properties of metal materials are primarily 
affected by the mobility of dislocations. 
By modulating the GSF energy profile with 
the introduction of different element atoms, 
the recent works by Pei etal. \cite{Pei2013, Pei2015} have reported the improvement 
of the mechanical properties of metals (such as Mg, Al). 
Hexagonal close-packed (hcp) zirconium and its alloys are used widely as structural 
materials in nuclear reactors due to their low neutron absorption cross section, 
good mechanical properties and superior corrosion resistance \cite{Zhilyaev2009}. 
Particularly, they are used as the cladding materials in light water nuclear reactors. 
The effects of H atoms in Zr and its alloys are a long term issue 
with respect to the reducement of their mechanical performances. 
The earlier works announced that the addition of H atoms would drastically decrease 
the GSF energy of the typical slip planes (primary basal plane and prismatic plane) 
in hcp Zr \cite{Legrand1984, Domain2002, Domain2004}. 
But the later literature by Udagawa et al. \cite{Udagawa2010} explained the main 
reason of hydrogen embrittlement of Zr alloys to be the brittle nature of the hydride 
rather than the solid solution of Zr-H system. 

Recently, Wu el al.\cite{Wu2010} proposed a  interpretation for the GSF features 
of different pure hcp metals based on the correlation of GSF and the unit 
cell volume as well as the ratio of lattice constant (c/a). 
Shang et al. \cite{Shang2014} tried to explain the 
physical mechanism of the GSF modification caused by alloying with other 
elements in hcp Mg by employing the analysis method of differential 
charge density. However, it is still unclear how H atoms interact with Zr atoms 
beside the slip planes, and the physical reasons of the reduction of the GSF 
energies in hcp Zr containing H atoms.
The purpose of this work is to systematically investigate the interaction 
of H atoms and the stacking faults in hcp Zr. The dependences of 
1/3$<10\bar{1}0>$\{0001\} and 1/3$<11\bar{2}0>$\{0001\} GSF 
energy profiles on the distance between H 
atoms and the slip plane as well as the H concentration  are calculated 
based on the density functional theory (DFT) method. 
Meanwhile, the Bader charge in combination with the structural features 
are adapted to explored the physical origin of reduction of GSF energy 
caused by H atoms neighboring the slip plane along two shift directions. 
The results demonstrate that the Coulomb interactions between 
positively charged Zr atoms beside the slip planes cause 
the decrease of energy barrier of 1/3$<10\bar{1}0>$\{0001\} 
GSF structure, whereas for 1/3$<11\bar{2}0>$\{0001\} GSF structure, 
the wide atomic space of \{0001\} planes weakens the Coulomb repulsion, 
and the energy barrier for the atomic plane shifting would be affected 
slightly. But as the increase of H concentration around the slip region, 
the enhanced Coulomb interaction between the slipping planes obviously reduces 
GSF energies of both GSF structures. 
Accordingly, the calculation method and model is given in Section II, followed by
the results and discussions including DFT calculation and Bader charge analysis. 
The conclusions will be presented finally.                                                                    

\section{Methods and Model}
\subsection{Computational method}
Our DFT calculations were performed using the projector augmented wave (PAW) \cite{Blochl1994} 
as implemented in the Vienna ab initio simulation package (VASP) \cite{Kresse1996}.
The Perdew-Burke-Ernzerhof (PBE) \cite{Perdew1996} scheme
of the generalized gradient approximation (GGA) is adapted to describe exchange and correlation.
To get accurate results, the plane wave cut-off energy is chosen as 400 eV for all calculation.
The Brillouin zone sampling employs the $\gamma$-centered grids of kpoints of 
11 $\times$ 11 $\times$ 1  
in terms of Monkhorst-Pack scheme \cite{Monkhorst1976}.
The quasi-Newton  algorithm was used to relax the ions until the energy and the 
Hellmann-Feyman force was less than $10^{-4}$ ev and 
0.05 eV/{\AA}.
\subsection{Theoretical model    }

\begin{figure}[!htbp]
\centering
\includegraphics[width=0.8\textwidth]{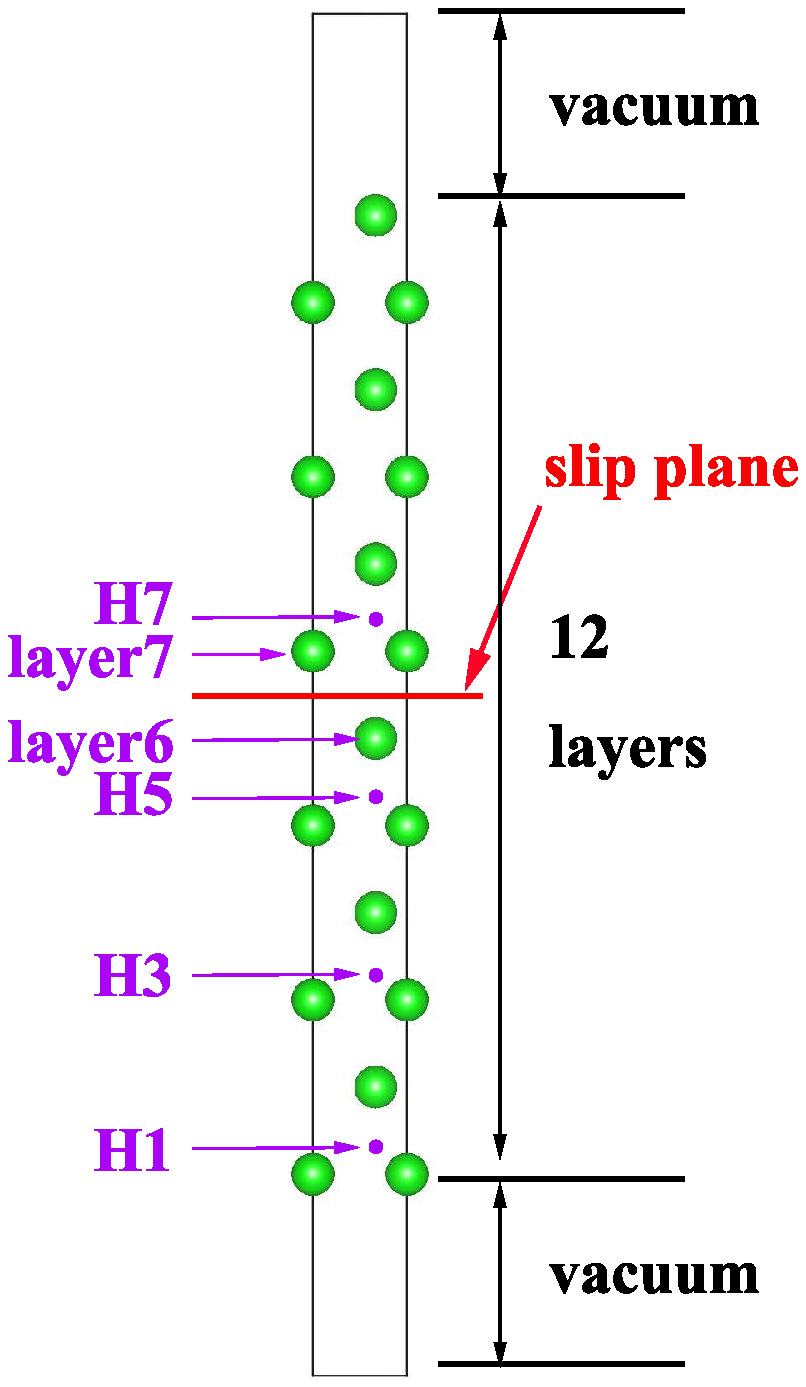}
\caption{The model of hcp Zr with H atom locating different position  
  represented by VESTA\cite{Momma2011}.
Zr: big green spheres; H: small purple spheres. 
H1, H3, H5 stand for pure 12 layers hcp Zr with
one H atom locating in the tetrahedron above layer1, layer3, layer5 separately. H57 presents that
two H atoms are putted into two tetrahedrons above layer5 and layer7 simultaneously.}
\label{structure}
\end{figure}

The volume and atom position are allowed to fully relaxed to obtain the lattice 
parameter of single hcp zirconium. Our calculated results (a = 3.235 {\AA}, c/a = 1.604) 
agrees well with experimental \cite{Jomard1998} (a = 3.23 {\AA}, c/a = 1.594) and other theoretical
results\cite{Udagawa2010}. 
Hydrogen as a small atom will locate tetrahedron or octahedron 
interstitial position when it gets into the hcp zirconium system. A 3$\times$ 3 $\times$ 2 supercell
containing 36 Zr atoms and 1 H atom was constructed to investigate which position H atom locate 
is stable. The total energy difference (denoted as E(T)-E(O)) between these two configurations we 
calculated is -0.052 ev , which is in good agreement with other theoretical 
observations (-0.057 ev \cite{Domain2002}, -0.061 ev \cite{Christensen2015}), 
indicating T-site is stable. 
Consequently, the model we adapted in this work was constructed with H atom in tetrahedron interstitial
position.
As illustrated in Fig. 1, a 1$\times$ 1 $\times$ 12 supercell with 8 {\AA} vacuum between them are 
constructed to prevent any  interactions between periodic images
for calculating GSF energy curves. The stable interstitial
position is T-site when a hydrogen atom is added to hcp zirconium system.
We named layer1, layer2, \ldots, layer12 from the bottom to the top of 
the hcp zirconium supercell. We called the structure as H1, H3, H5 separately,  
when we put a hydrogen atom in a tetrahedron above layer1, layer3, layer5. That is , the hydrogen 
is the closest to the slip plane in H5. Meanwhile, we place two hydrogen atoms in two tetrahedrons 
above layer5 and layer7 separately to study the influences of the hydrogen concentration on
the GSF enery,
which is denoted as H57.

\section{Results and discussion}
\subsection{GSF energy of basal plane                     }

\begin{figure}[!htbp]
\centering
\includegraphics[width=0.8\textwidth]{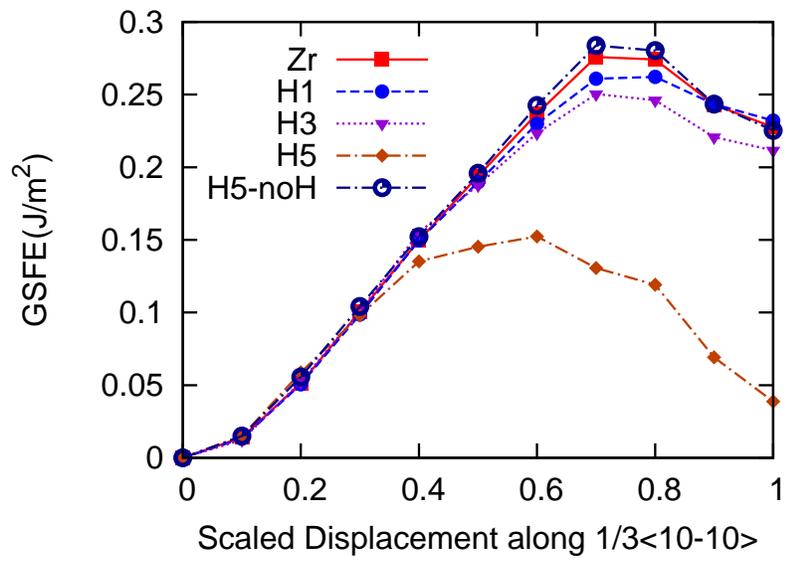}
\caption{The GSF energy curves of pure Zr and Zr-H system with H atom locating different position   
  versus fractional displacement  along $<10\bar{1}0>$ direction in the basal plane}
\label{GSFE}
\end{figure}

\begin{figure}[!htbp]
\centering
\includegraphics[width=0.8\textwidth]{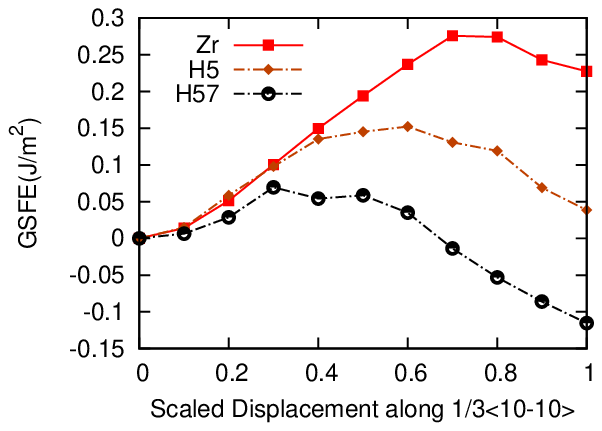}
\caption{The GSF energy curves of pure Zr and Zr-H system with different concentration of  H atom 
  versus fractional displacement  along $<10\bar{1}0>$ direction in the basal plane}
\label{structure}
\end{figure}

\begin{figure}[!htbp]
\centering
\includegraphics[width=0.8\textwidth]{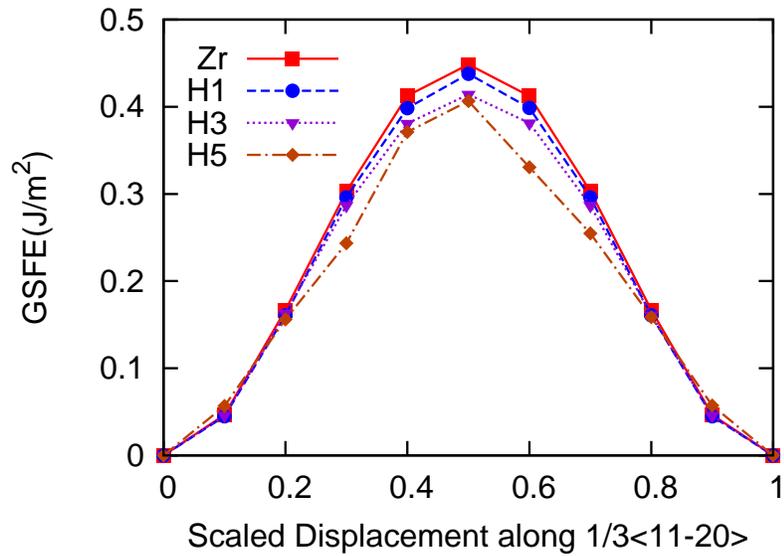}
\caption{GSF energy curves
 versus fractional displacement  along $<11\bar{2}0>$ direction in the basal plane} 
\label{structure}
\end{figure}

\begin{figure}[!htbp]
\centering
\includegraphics[width=0.8\textwidth]{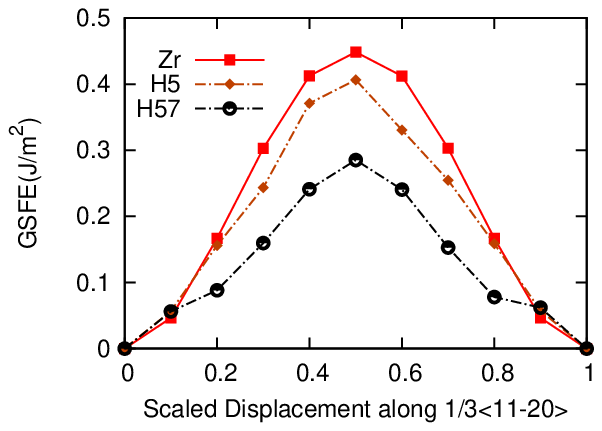}
\caption{The GSF energy curves of pure Zr and Zr-H system with different concentration of  H atom 
  versus fractional displacement  along $<11\bar{2}0>$ direction in the basal plane
} 
\label{structure}
\end{figure}

Fig.2 presents the GSF energy curves of \{0001\} along $<10\bar{1}0>$ as a function of fractional 
displacement for pure Zr and the Zr-H system. For the pure Zr, the maximum value 
(about 0.276 $\rm{J/m^{2}}$) as well as the fault position it reached (at
7/10, 8/10 of total displacement) agree well with previous results (about 
0.275 $\rm{J/m^{2}}$)\cite{Udagawa2010},
 confirming the reliability of our calculations.

Fig.2 shows the remote H atom from the slip plane (H1,H3) slightly decrease the GSF energy curves of the 
hcp Zr , while the nearest H atom (H5) drastically decrease the 
GSF energy curve.  In the H5 case, the $\gamma_{US}$ (unstable stacking energy) 
and $\gamma_{SF}$ (stacking fault energy)  are predicted to decline from 
0.276 $\rm{J/m^{2}}$ and 0.228 $\rm{J/m^{2}}$ to 0.152 $\rm{J/m^{2}}$ and 0.039 
$\rm{J/m^{2}}$ (nearly 45\% and 83\%) 
due to the addition of the H atom nearest to the slip plane.
According to the study of Wu el al.\cite{Wu2010}, the structural properties 
are important factors that influence the GSF energy curves for pure metals.
Therefore, we suppose the dramatically change of GSF energy curve is because of volume expansion. 
So we remove the H atom from the relaxed H5 case to study the effect of volume expansion
on the GSF energy
curve, and the result is shown as H5-noH in Fig.2. However, It is found that the $\gamma_{US}$ 
is slightly increased
because layer6 and layer7 are closer due to the squeezing force 
of H atom to its neighboring Zr atoms. 
So structural characters are not enough to fully understand the trends of GSF energy curve after adding
a H atom.

In order to further study the effects of the hydrogen concentration 
on the GSF energy curve, we calculated
the GSF energy curve
of H57 , which is plotted in Fig. 3 as well as the ones of Zr and H5. 
It is clearly found that the $\gamma_{US}$
was rapidly reduced to 0.070 $\rm{J/m^{2}}$ for H57 from 0.152 $\rm{J/m^{2}}$ for H5 
after the addition of one more H atom. Meanwhile, the $\gamma_{SF}$ of H5 
dropped from 0.039 $\rm{J/m^{2}}$
to -0.115 $\rm{J/m^{2}}$ for H57. It means that the stable stack fault of 
H57 structure is  more stable than the unchanged one.

\begin{table*}[htbp]
\scriptsize
\caption{ Influence of H on the GSF energy curve ($\rm{J/m^{2}}$) in Zr for the basal plane     
}
\begin{center}
  \resizebox{\textwidth}{!}{
  \begin{tabular}{cccc} 
  \hline
  &$\gamma_{US}<10\bar{1}0>$&$\gamma_{SF}<10\bar{1}0>$&$\gamma_{US}<11\bar{2}0>$\\ \hline
pure Zr&0.276&0.227&0.449\\
H1&0.262&0.232&0.438\\
H3&0.250&0.212&0.414\\
H5&0.152&0.039&0.407\\
H57&0.070&-0.115&0.285\\
\hline
\end{tabular}
}
\end{center}
\label{Table.1}
\end{table*}

Similar to Fig.2 and Fig.3, we plotted out the results along $<11\bar{2}0>$ of \{0001\} in 
Fig.4 and Fig.5, which is
another common slip pattern in basal plane for hcp metals.
Like the results along $<11\bar{2}0>$, the $\gamma_{US}$ is becoming smaller as the H atom 
closer to the slip plane or increasing the content of H atom. The details are shown in Table. 1.

However, it is observed that the magnitude of decline along $<11\bar{2}0>$ direction is much smaller
than that along $<10\bar{1}0>$ direction for the structures containing one H atom.
For the case of H5, the $\gamma_{US}$ along $<11\bar{2}0>$ is merely 0.042 $\rm{J/m^{2}}$ smaller 
than the $\gamma_{US}$ in pure
Zr.
The reasons behind this will be discussed carefully in next section considering both
structural and electronic properties.

\subsection{Bader charge analyse}

\begin{figure}[!htbp]
\centering
\includegraphics[width=0.8\textwidth]{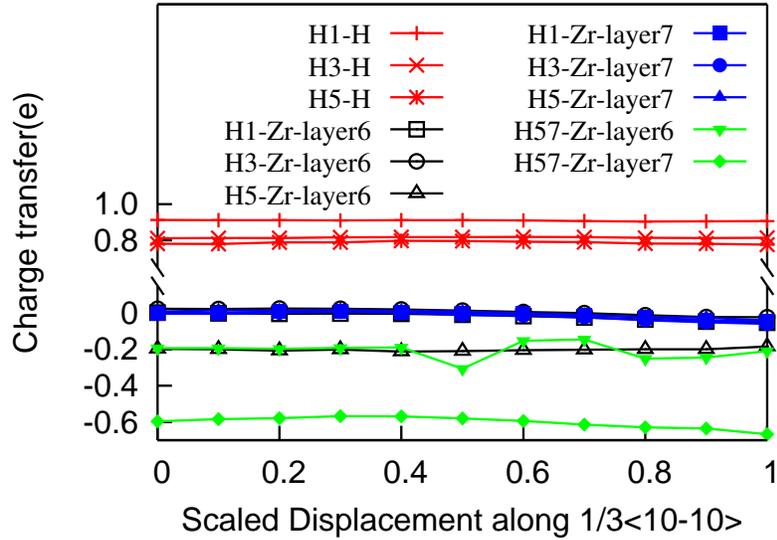}
\caption{The bader charges transfer of H and Zr atom of layer6 and layer7 in H1, H3, H5, H57 
structure versus fractional displacement  along $<10\bar{1}0>$ direction in the basal plane.
The positive values present obtaining electrons while the negative ones mean losing electrons.} 
\label{structure}
\end{figure}

\begin{table*}[htbp]
\scriptsize
\caption{ The numbers of charge transfer caused by H, C, N in H5 structre.
The positive values present obtaining electrons while the negative ones mean losing electrons. 
}
\begin{center}
  \resizebox{\textwidth}{!}{
  \begin{tabular}{cccc} 
  \hline
  Charge transfer(e)&$\rm{Zr_{12}H}$&$\rm{Zr_{12}C}$&$\rm{Zr_{12}N}$\\ \hline
solute atom&0.79&1.77&1.71\\
Zr(layer6)&-0.21&-0.49&-0.48\\
\hline
\end{tabular}
}
\end{center}
\label{Table.5}
\end{table*}

\begin{figure}[!htbp]
\centering
\includegraphics[width=0.8\textwidth]{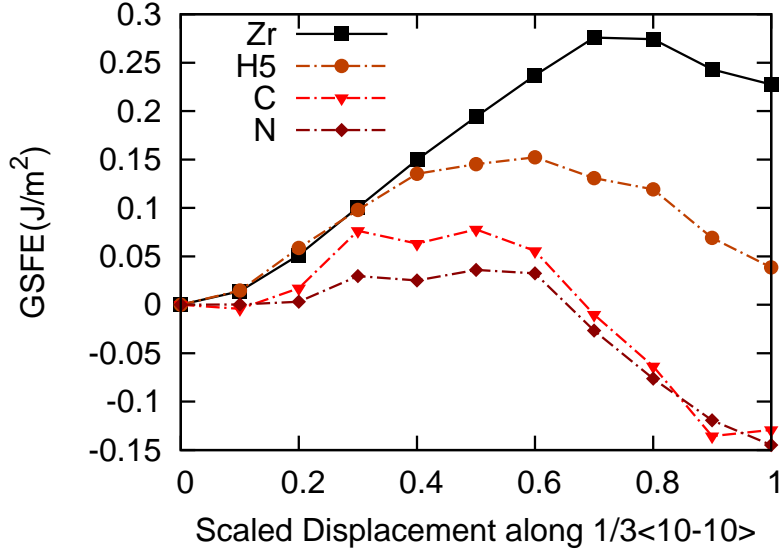}
\caption{The effect of addition of H, C, N on the GSF energies in H5 
structure versus fractional displacement  along $<10\bar{1}0>$ direction in the basal plane} 
\label{structure}
\end{figure}

It's natural to understand the change of $\gamma_{US}$ in 
the view of charge properties.
Actually, in a recently work\cite{Shang2014}, the authors attribute the variations 
of $\gamma_{US}$ caused by the adding of alloy elements to the redistribution of charge
density of Mg-based alloys. 
As for our case, we calculated the charge transfers of each atom caused
by the addition of H atom along $<10\bar{1}0>$ in terms of Bader charge analysis\cite{Henkelman2006,
Sanville2007},
which are presented in Fig. 4 as a function of displacement .
It is found that: (1) the H atom obtain electrons showing stronger electronegativity 
while the Zr atom lose electrons. (2) the values of charge transfer for both 
H and Zr in H1, H3, H5, H57 structures 
remain nearly unchanged in the whole shift process. 
(3) the charge transfers of Zr 
in layer6 are almost zero due to being far away from the H atom in H1, H3 structures, 
while in the H5 and H57 cases, the Zr atom in layer6 lose about 0.2 electrons.
Furthermore, the Zr atom in layer7 of H57 lose as many as about 0.6 electrons.

There will be a repulsive force between two atoms with the same kind of charge.
Therefore it is natural to propose that the Coulomb repulsion is the reason 
behind the decline phenomenons of GSF energy curves caused by H atom in hcp Zr. 
The Coulomb repulsion between the Zr atoms in layer6 and layer7 will become stronger 
due to  the increase  of charges of Zr in layer6.
The relationship is more obvious for the H57 case, where the $\gamma_{US}$ is further reduced compared
to the H5 case because of more charge transfers for Zr atoms in layer7.

That is to say, there maybe a direct relationship between the magnitude of decline    
and the amount of charge transfer caused by the H atom.
In order to test and clarify the rationality of our hypothesis, we replaced H atom by C, N separately
in H5 structure, and then
their GSF energy curves along $<10\bar{1}0>$ and charge transfers were calculated,
obtained results shown in Fig. 7 and Table.2.

From Fig.7, We find that the magnitudes of decline induced by C, N  are much bigger than
the ones caused by H, consistent with the results that the charge transfers caused by C, N
are much more than the ones caused by H, as shown in Table 2.  
Furthermore, it is reasonable to conclude that the similar GSF energy curves of Zr-C, Zr-N system
result from the nearly same charge transfers caused by C, N separately.
As a result, our proposal are confirmed
directly in basal plane of hcp Zr.

As mentioned in last section, problems are still existed in the results of GSF energy 
along $<11\bar{2}0>$
since the magnitude of decline is much more smaller than that along $<10\bar{1}0>$. 
With the H5 structure as an example, it is found that the distance between the 
layer6 and layer7 along $<11\bar{2}0>$ is about 0.15 {\AA} larger than the one 
along $<10\bar{1}0>$ in relaxed H5 structure.
It will result in weakening the influence of the charge transfer on the GSF energy since
0.15 {\AA} is considerable in the atom scale. Maybe the Coulomb repulsion could be 
negligible when the distance reaches certain value.

Consequently, the phenomenons of decline caused by H atom in hcp Zr are interpreted 
qualitatively  
by charge transfers and structure properties.

\section{Conclusion}

In present work, the GSF energy curves of \{0001\} along $<10\bar{1}0>$ and 
$<11\bar{2}0>$  directions for both pure Zr 
and four Zr-H systems were calculated using Density Functional Theory 
as implemented in VASP. The effects of distance to the slip plane and concentration of H atom on 
the GSF energy curves were thoroughly investigated. In order to clarify the reason of the declines of 
GSF energy curves caused by H atom, the method of Bader charge analysis is adapted. 
Our results are summarized as follows:
(1) The decreasing effect become
more obviously as the H atom closer to the slip plane. Specially, in the H5 case
along $<10\bar{1}0>$, the $\gamma_{US}$ 
and $\gamma_{SF}$ are reduced significantly by 45\% and 83\% separately. 
(2) The content of H atom
is another factor that decrease the GSF energy profile.
(3) Using the Bader charge analysis, we find there are more charge transfers in layer6 when the H 
atom is closer to the slip plane. More charge transfers mean stronger Coulomb repulsion, and  that
will lead to larger magnitude of decline in GSF energy curves.
(4) The magnitude of decline caused by H atom along $<11\bar{2}0>$ direction is much smaller
than that along $<10\bar{1}0>$ direction for the structures containing one H atom.
Coulomb repulsion will be weakened because  the  distance between the 
layer6 and layer7 along $<11\bar{2}0>$ is about 0.15 {\AA} larger than the one 
along $<10\bar{1}0>$ in relaxed H5 structure. 
Therefore, a less obvious decrease of GSF energy curve (Fig.5) along $<11\bar{2}0>$ is reasonable.
(5) Our conclusions are verified by replacing H with C and N in H5 structure. 

This work provides more deeper understandings on the behavior of H in 
hcp Zr, and the result that the dislocation occur more easier in the part there are 
more H atoms is predicted,
which will be useful for the researchers working in the light water nuclear field.

\newpage
\noindent \textbf{Acknowledgments}\\

\noindent This work was supported by the National Science Foundation of China under Grant Nos. 1127522a
\& NSAF U1230202, special Funds for Major State Basic Research Project of China (973) under Grant
No. 2012CB933702, Hefei Center for Physical Science and Technology under Grant No. 2012FXZY004,
Anhui Provincial Natural Science Foundation under Grant No. 1208085QA05, and Director Grants of
CASHIPS. Part of the calculations were performed at the Center for Computational Science of CASHIPS,
the ScGrid of Supercomputing Center, and the Computer Network Information Center of the Chinese Academy of
Sciences.

\bibliographystyle{elsarticle-num}
\bibliography{Refs}
\newpage

\end{document}